# Machine Learning Potential for Modelling $H_2$ Adsorption/Diffusion in MOF with Open Metal Sites

Shanping Liu, Romain Dupuis, Dong Fan, Salma Benzaria, Michael Bonneau, Prashant Bhatt, Mohamed Eddaoudi, and Guillaume Maurin*

**ABSTRACT:** Metal-organic frameworks (MOFs) incorporating open metal sites (OMS) have been identified as promising sorbents for many societally relevant-adsorption applications including $CO_2$ capture, natural gas purification and $H_2$ storage. This has been ascribed to strong specific interactions between OMS and the guest molecules that enable to achieve an effective capture even at low gas pressure conditions. In particular, the presence OMS in MOFs was demonstrated to substantially boost the $H_2$ binding energy for achieving high adsorbed hydrogen densities and large usable hydrogen capacities. So far, there is a critical bottleneck to computationally gain a complete understanding of the thermodynamics and dynamics of $H_2$ in this sub-class of MOFs since the generic interatomic potentials (IPs) fail to accurately capture the OMS/$H_2$ interactions. This clearly hampers the computational-assisted identification of existing or novel MOFs since the standard high-throughput screening approach based on generic IPs are not applicable. Therefore, there is a critical need to derive IPs to achieve accurate and effective evaluation of MOFs for $H_2$ adsorption. On this path, as a proof-of-concept, the Al-soc-MOF containing Al-OMS, previously envisaged as a potential candidate for $H_2$ adsorption, was selected and a machine learning potential (MLP) was derived from a dataset initially generated by *ab-initio* molecular dynamics (AIMD) simulations. This MLP was further implemented in MD simulations to explore the binding modes of $H_2$ as well as its temperature dependence distribution in the MOFs pores from 10K to 90K. MLP- Grand Canonical Monte Carlo (GCMC) simulations were further performed to predict the $H_2$ sorption isotherm of Al-soc-MOF at 77K that was further confirmed by gravimetric sorption measurements. As a further step, MLP-based MD simulations were conducted to anticipate the kinetics of $H_2$ in this MOF. This work delivers the first MLP able to describe accurately the interactions between the challenging $H_2$ guest molecule and MOFs containing OMS. This innovative strategy applied to one of the most complex molecules owing to its highly polarizable nature alongside its quantum-mechanical effects that are only accurately described by quantum calculations, paves the way towards a more systematic accurate and efficient *in silico* assessment of the MOFs containing OMS for $H_2$ adsorption and beyond to the low-pressure capture/sensing of diverse molecules.

## INTRODUCTION

Metal-Organic Frameworks (MOFs) constructed from metal ions/metal clusters connected to organic linkers have been widely studied and continue to gain momentum as a class of porous frameworks.[1-5] Their unique tunability in terms of chemical functionality, architecture and pore size/shape make them potentially applicable in many fields, including gas capture/storage, separation, catalysis, biomedicine and sensing among others.[6-8] One sub-class of MOFs contains open metal sites (OMS) also called coordinatively unsaturated sites (CUS) on the cluster nodes, to which guest molecules can readily bind.[9-11] The formation of this strong metal-molecules bond has been demonstrated to play a key role for not only initiating a myriad of catalytic reaction in the MOF pores but also for selectively adsorbing a desired molecule.[11] Typically, $Mg(II)_2$(dobpdc) with its pore wall decorated by Mg-OMS is one of the prototypical MOFs for the selective capture of $CO_2$ over a range of other molecules ($N_2$, $CH_4$, $H_2O$....),[12] while MIL-100(CrIII) owing to its Cr-OMS was demonstrated to be highly selective for $N_2$ over $CH_4$ of great interest for natural gas purification.[13] MOFs incorporating OMS have also shown promises for the storage/delivery of hydrogen ($H_2$),[14-16] a highly relevant energy vector, especially as a replacement of traditional fossil fuels. This societally-relevant topic is of importance since Net-zero hydrogen with a GreenHouse Gas (GHG) footprint of zero is expected to provide up to 24% of the total EU energy demand.[17] Typically, $H_2$ storage currently involves the use of high pressure and/or cryogenic temperatures that implies high additional costs and safety issues.[18] The challenge in this field is to identify porous materials able to adsorb reversibly high $H_2$ uptake and concurrently maximize deliverable $H_2$ capacity. Since the first coordinately saturated MOF tested for $H_2$ adsorption,[19] i.e., MOF-5, the MOF community designed a series of porous materials incorporating OMS over the last two decades with the objective to substantially enhance $H_2$ binding energy/adsorption enthalpy for achieving high adsorbed hydrogen densities. Typically, $V_2Cl_{2.8}$(btdd) containing a high concentration of V(II) sites demonstrated high usable hydrogen capacities that exceed that of compressed storage under the same operating conditions.[20]

This list of examples highlights the key role played by OMS in many adsorption-related properties of MOFs and the need to gain an in-depth understanding of the OMS-guest interactions towards the refinement of MOFs with improved performances. Molecular simulation has proven to be a complementary tool to sophisticated characterization techniques to precisely characterize the interactions between guest molecules and MOFs containing OMS. To effectively simulate these rather complex host/guest systems, a reliable interatomic potential is required to accurately describe the potential energy surface (PES). Here, the challenge lies in a correct description of the weaker intermolecular forces between the guest molecules and the host framework, e.g., coulombic and van der Waals interactions as compared to the strong intramolecular forces within the host framework, e.g., metal ion-linker bonding. Quantum *ab-initio* approach, such as the dispersion-corrected density functional theory (DFT) provides precise determination of such interaction forces, however, its applicability is restricted to small systems containing less than hundreds of atoms due to its prohibitive computational cost. Additionally, *Ab initio* Molecular Dynamics (AIMD) simulations are also limited to pico-second timescales. These shortcomings make *ab-initio* approach inappropriate to explore the guest adsorption in MOFs at long-time and large-length scales. Classical force fields (also called Interatomic Potentials (IPs)) simulations offer an alternative approach that has been widely employed to explore the adsorption of guest molecules in MOFs at larger scale.[21-23] The large majority of these reported theoretical works relied on the application of Lorentz Berthelot mixing rules between generic IPs, e.g., universal force field (UFF) and Dreiding among others for the MOF framework,[24,25] and diverse IP models for the guests. Although this simplified approach has been shown to capture quite well the interactions between small guest molecules and coordinatively saturated MOFs, it cannot anymore be applicable to MOFs containing OMS that induce high polarization in the adsorbed molecules.[26] This statement hampers the computational-assisted identification of existing or novel MOFs since the standard high-throughput screening approach based on generic IPs are not applicable. Indeed such MOF-OMS/guest molecule interaction requires a specific IP parameterization that is far to be a trivial task.[27] Therefore, there is a critical need to move beyond classical approaches and derive IPs combining high efficiency and high accuracy, capable of describing the overall interactions that are in play in MOF-OMS/guest molecule systems. One promising approach to achieve this objective is the development of machine learning potentials (MLPs), which are trained on database preliminary generated by DFT calculations. Development of MLPs for describing PES in condensed matter was pioneered by Behler and Parrinello.[28] This was achieved by considering physically meaningful descriptors which represent very well the atomic structure. By directly fitting the relationships between structure and energy, MLPs generally enables to reproduce complex interactions more accurately than classical IP and more efficiency (less expensive computational cost) than DFT. MLPs applied to MOFs is therefore expected to gain an unprecedented description of the MOF-OMS/guest molecule systems with high-accuracy accounting for the overall forces present in this complex system. So far, the development of MLPs for MOFs has been limited to only a very few cases. Behler *et al*., first derived a high-dimensional neural network MLP to effectively describe the crystal structure of MOF-5.[29] Johnson *et al*. further combined MLP for the UiO-66 framework with classic IPs for rare gases to explore the host/guest interactions.[30] Very recently, Vandenhaute *et al*. built a neural network potential with parallelized sampling and on-the-fly training to explore the phase transformation for different MOFs.[31] Zheng *et al*. implemented a novel MLP to explore the $CO_2$ binding mode and diffusion in MOF-74 containing Mg(II) -OMS.[32] Similarly, Shaidu *et al*. developed a neural network MILP for the exploration of $CO_2$ binding in amine-appended $Mg_2$(dobpdc).[33] This preliminary study highlights that MLP offers a great opportunity to explore the most complex MOF host-guest interactions at large scale.

Achieving an accurate description of the interactions between $H_2$ and any host frameworks at the force field level is by far more challenging since quantum-mechanical effects become significant for this smallest molecule particularly at cryogenic temperature and its polarizability in confined space plays also a key role.[34-36] Many force fields are available in the literature for describing $H_2$, including the most sophisticated ones that implements the Feynman-Hibbs variational approach to take into account the quantum effects.[37] However once combined with generic force fields to describe coordinatively saturated MOF frameworks, they frequently failed to reproduce the thermodynamics/dynamics of $H_2$@MOF systems that motivated tedious force field re-parametrization based on experimental data for some specific systems.[38,39] The complexity becomes even higher when one deals with the interactions between $H_2$ and MOF-OMS that have been solely treated so far at the pure quantum-level.[40]

Herein, we aim to develop a MLP using a deep neural network to accurately explore the adsorption and diffusion behaviors of $H_2$ in a prototypical MOF containing OMS. As a proof-of-concept, the soc-MOF platform composed of 6-connected metal trinuclear molecular building block and a 4-connected rectangular-planar organic ligand, 3,3′,5,5′-azobenzenetetracarboxylate linker was selected owing to its key features, i.e., high OMS concentration and optimal pore size of about 1 nm and geometry, that were demonstrated favorable for an efficient $H_2$ adsorption.[41,42] Molecular Dynamics simulations implementing MLP initially trained from a series of configurations generated by AIMD at 77 K enabled to accurately capture the binding modes of $H_2$ towards the OMS and beyond elucidate the $H_2$ distribution in the overall porosity of the soc-MOF in a wide range of temperature spanning from 10K to 88 K.

MLP-grand canonical Monte Carlo (GCMC) simulations further predicted the $H_2$ adsorption isotherm at 77 K in the low domain of pressure that was validated by a good agreement with the gravimetric adsorption isotherm freshly collected on a well-activated soc-MOF sample. MLP-MD simulations finally delivered a microscopic picture of the diffusion of $H_2$ in this soc-MOF. This computational work delivers the first MLP able to accurately describe the interactions between $H_2$ and MOF-OMS, a key to gain an in-depth understanding of the $H_2$/MOF-OMS interactions of importance to further develop advanced MOF for efficient $H_2$ storage.

## RESULTS AND DISCUSSION

### Overall Computational Strategy

Figure 1 illustrates the overall workflow we implemented to develop and validate a MLP to accurately describe the host-guest interactions in the prototypical Al-soc-MOF/$H_2$ system. The first step (Figure 1a) involved the construction of a reliable atomistic guest-loaded model. One unit cell of the cage-like Al-soc MOF,[43] containing metal Al oxo-trimer nodes with 1 metal bounded to the OH counter-ions and 2 OMS, was initially loaded by a representative $H_2$ uptake (24 molecules, i.e., 0.92wt%) that mimics the low-pressure adsorption regime of the analogue In-soc-MOF previously assessed by gravimetry measurements (0.15 bar at 77 K).[44-46] Such a scenario was selected since the first adsorbed $H_2$ molecules are expected to interact preferentially with the OMS and indeed offer an optimum scenario to derive the target Al-OMS/$H_2$ MLP.

The construction of the MLP then involves three steps: generation of high quality quantum training data, creation of effective descriptors by fingerprinting the local atomic environment, and establishment of a robust mapping between the descriptors and atomic energies and forces. The training set on the Al-soc-MOF/$H_2$ models were thus generated (Figure 1b) by Ab initio Molecular Dynamics (AIMD) at 77 K for 7 ps, accumulating 11k+ data points to capture configurations, energies, and forces. To ensure a quality of the training data, we performed a pre-equalization simulation of ~1.4 ps before the production runs. The DeepMD-kit program was employed to build the descriptors based on the produced training data.[47, 48] Within DeepMD-kit, a continuous neural network was utilized to establish a robust mapping between the effective descriptors and atomic energies to generate the MLP (Figure 1c). Concurrently, the accuracy of the fitted MLP was validated using the separate validation dataset (use of 782 data points), assessing the performance and reliability of the MLP. The next step consisted of implementing the derived MLP in classical molecular dynamics (MD) simulations to externally validate its reliability to accurately describe the preferential location/distribution of $H_2$ in the Al-soc-MOF and importantly the $H_2$/Al-OMS interactions (Figure 1d). Finally, the robustness of this MLP was challenged by running (i) Monte Carlo (MC) simulations to predict the $H_2$ adsorption isotherm of Al-soc-MOF in the low-pressure region that was further compared to corresponding experimental data and MD simulations to assess the $H_2$ transport (Figure 1e).

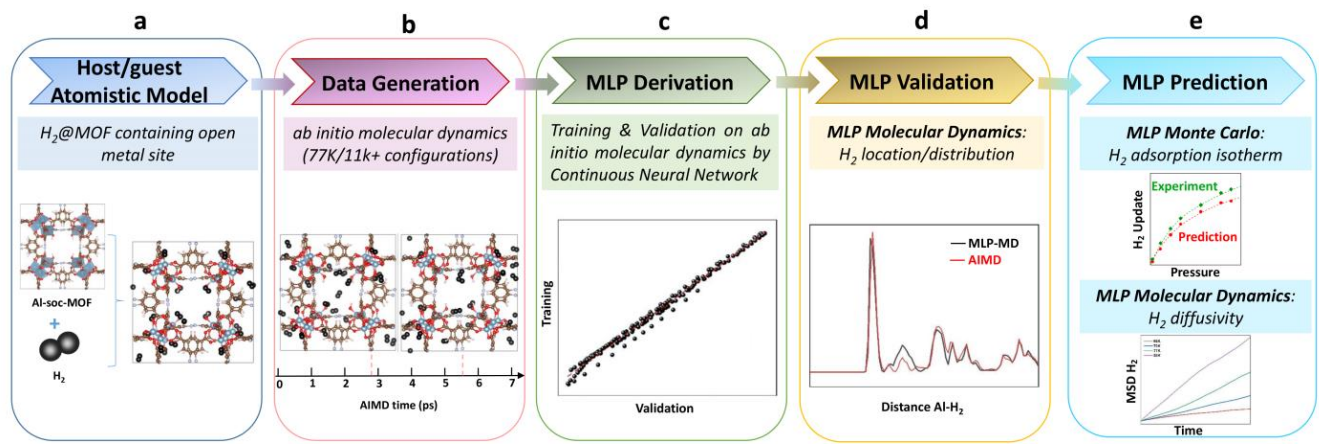

**Figure 1.** Workflow devised for the derivation and validation of a machine learning potential for $H_2$@Al-soc-MOF towards prediction. (a) Construction of the atomistic model for the host/guest system. (b) Generation of the training datasets using *ab Initio* Molecular Dynamics simulations. (c) Training and validation of the MLP using continuous neural network. (d) External validation of the trained MLP with its implementation in Molecular Dynamics simulations and (e) MLP-Molecular Dynamics and MLP-Mont Carlo Predictions.

## MLP Derivation by neural network algorithm

The training process in the DeepMD-kit program aims to optimize the parameters in the filter and fitting neural network by minimizing the loss function, also called learning rate. This loss function is calculated based on the root mean square errors (RMSEs) of energies and forces when compared to the AIMD simulated data. We first checked the reliability of the training set generated at 77 K for the 24$H_2$@Al-soc-MOF model system illustrated in Figure 2a. Figure 2b evidences a very tiny temperature fluctuation over the AIMD run and the total energy of all generated configurations shows a maximum deviation of 0.2% (Figure 2b). This observation emphasizes that the AIMD data are well equilibrated and offer an ideal platform for the MLP derivation. Figure 2c reports that the loss function exponentially decreases with the increase of training loops. We froze the potential model when the loss function reaches $10^{-6}$, typically after approximately 500.000 training loops with a RMSE lower than 0.40 meV/atom (Figure 2d). To assess the accuracy of the so-trained MLP, we randomly selected 200 configurations from the training dataset, and the associated RMSE was 0.30 meV/atom (Figure 2e). We further randomly choose 200 configurations from the validation dataset leading to identical RMSE of 0.29 meV/atom (Figure 2e). All these RMSE values closely approach the convergence energy criteria generally applied in quantum calculations.[49] Additionally, the linear correlation between the MLP-predicted energies and AIMD-calculated energies is excellent, with R-squared values exceeding 0.99 for training and validation tests. Therefore, the MLP demonstrates the ability to accurately predict energies across the entire energy range of the training and validation dataset, achieving satisfactory accuracy when compared to quantum calculations.

## MLP Validation throughout MD simulations

MD simulations implementing the derived MLP were further performed to explore the average distribution of $H_2$ in the pores of Al-soc-MOF. Figure 3a reports the temperature fluctuation of the system over a 3 ns-long NVT MD simulations conducted first at very low temperature (10K), employing a Nosé-Hoover thermostat with a timestep of 0.5 fs. Illustrative snapshots selected over the AIMD trajectory show that $H_2$ molecules remain mostly located next to the MOF pore wall, privileging interactions with the inorganic nodes (Figure 3b). Analysis of the radial distribution function (RDF) calculated for the Al-OMS/$H_2$ pair reveals a preferential interaction between the guest molecule and the Al-OMS with an average separating distance of ~2.73 Å that aligns with that derived from the AIMD simulations conducted at the same temperature for 6 ps. Interestingly, this geometric adsorption feature is in excellent agreement with the conclusion drawn from the Density Functional Theory (DFT)-geometry optimized $H_2$@Al-soc-MOF structure at 0 K, associated with Al-OMS/$H_2$ distance of 2.71 Å reported for comparison in Figure 3c. Notably, the overall MLP- and AIMD-derived RDF profile for Al-OMS/$H_2$ match well (Figure 3c). Indeed apart from the reproduction of the first sharp and intense RDF peak, MLP-MD simulations enable to capture the low intensity peaks at 3.4 and 4.1 Å corresponding to the interactions between $H_2$ and the two other Al atoms present in the same oxo-centered trimer as well as additional peaks within (i) 5 to 7 Å range assigned to the interactions between $H_2$ and the Al atoms present in the neighboring oxo-centered trimers and at (ii) about 8.5-9.5 Å associated with the distance between $H_2$ and the most distant oxo-centered trimer in the periodic unit cell. This overall observation demonstrates unambiguously that the trained MLP successfully capture the quantum-derived structuring of $H_2$ in the MOF pores as shown by an excellent reproduction of all peak positions, widths and heights of the RDF plot for the Al-OMS/$H_2$ pair. Interestingly these findings demonstrate the robustness of the MLP derived at 77 K that remains valid when applied at lower temperature (10 K). Complementary MLP MD simulations were performed to examine the transferability of MLP to a wider temperature range from 10 to 88K. Examinations of the temperature evolution of the MD runs and the configurations show that the integrity of the MOF framework is maintained and the $H_2$@Al-soc-MOF interactions are well described within the overall temperature range (Figure S1). Analysis of the RDF plotted for the Al-OMS/$H_2$ pair at different temperatures (Figure 4d) shows that the intensity of the first main peak decreases as the temperature increases accompanied by a tiny shift towards longer separating distance. This observation is in line with an increased mobility of $H_2$ within the MOF pores as temperature rises. This highlights that at lower temperature $H_2$ molecules are much more localized next to the Al-OMS, which are the primary adsorption sites. At higher temperature $H_2$ becomes more mobile and tend to distribute at longer distance to the Al-OMS. This behavior is in line with a relatively moderate binding energy that prevents to maintain a strong coordination of H$_2$ towards the metal site once the thermal vibration increases. The RDF plot for the H$_2$-H$_2$ pair equally shows a broadening of the peak when the temperature increases as depicted in Figure S2 consistent with less localized/more mobile guest molecules at higher temperature. Furthermore, static analysis of the adsorption positions of H$_2$ molecules validates the reliability of our MLP in accurately describing the H$_2$ adsorption properties at various temperatures. This confirms the potential's ability to capture the temperature-dependent behavior of H$_2$ within the MOF system.

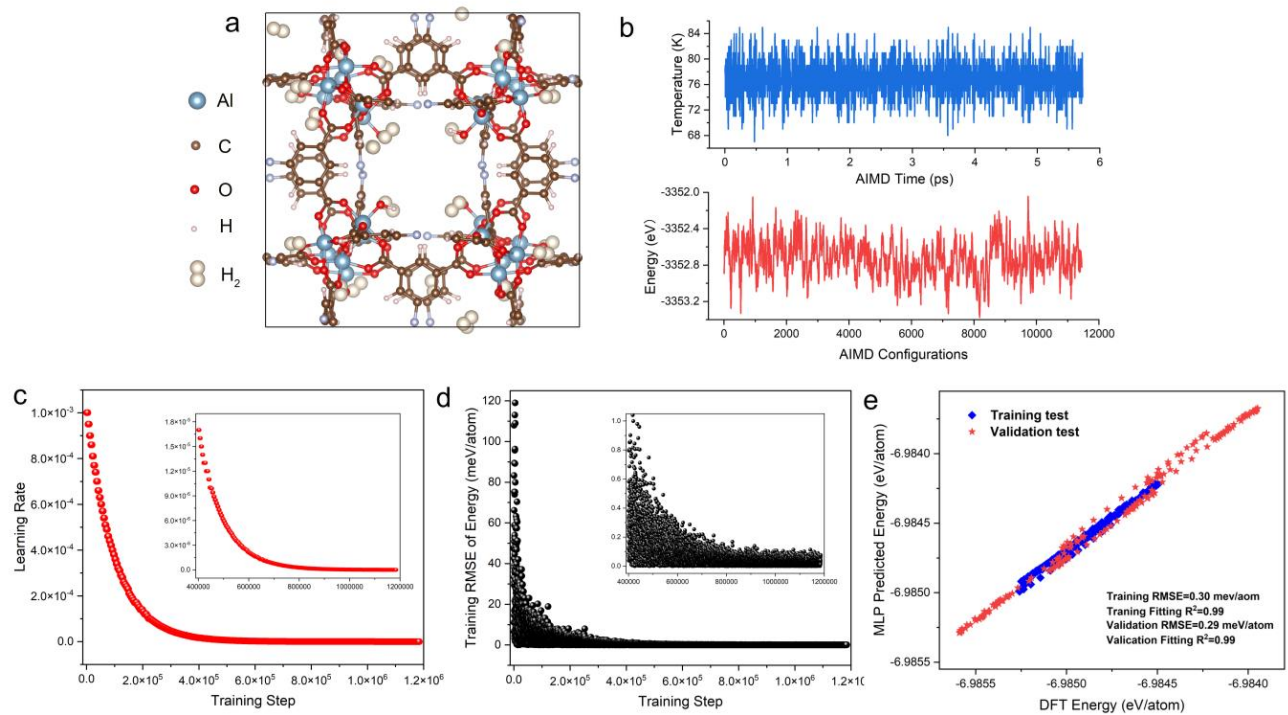

**Figure 2. MLP Derivation for $H_2$@Al-soc-MOF.** (a) Illustration of the 24$H_2$@Al-soc-MOF host-guest model system selected for the generation of the training/validation data points by AIMD simulations. (b) Reliability of the training data followed by plotting the temperature evolution over the AIMD trajectory and the energy fluctuation over the generated configurations. (c) Variation of the learning rate over the training loops. (d) Training RMSE of energy changes over the training loops. (e) Training and validation test with randomly selected data from the training and validation datasets.

This structural analysis highlights that the derived MLP achieves an accurate description of the overall H$_2$/Al-soc-MOF interactions in a wide range of temperature since it enables to finely capture the temperature-dependent location of the guest in this MOF. Beyond this observation, the derivation of such robust MLP paves the way towards the exploration of such complex host/guest system at longer timer scale (ns vs ps scale) and lower computational cost (about 300 times faster) compared to AIMD simulations (Table S1). This is of key importance particularly for probing the guest diffusion in MOF that operates at longer length and time scales than that accessible by AIMD simulations.

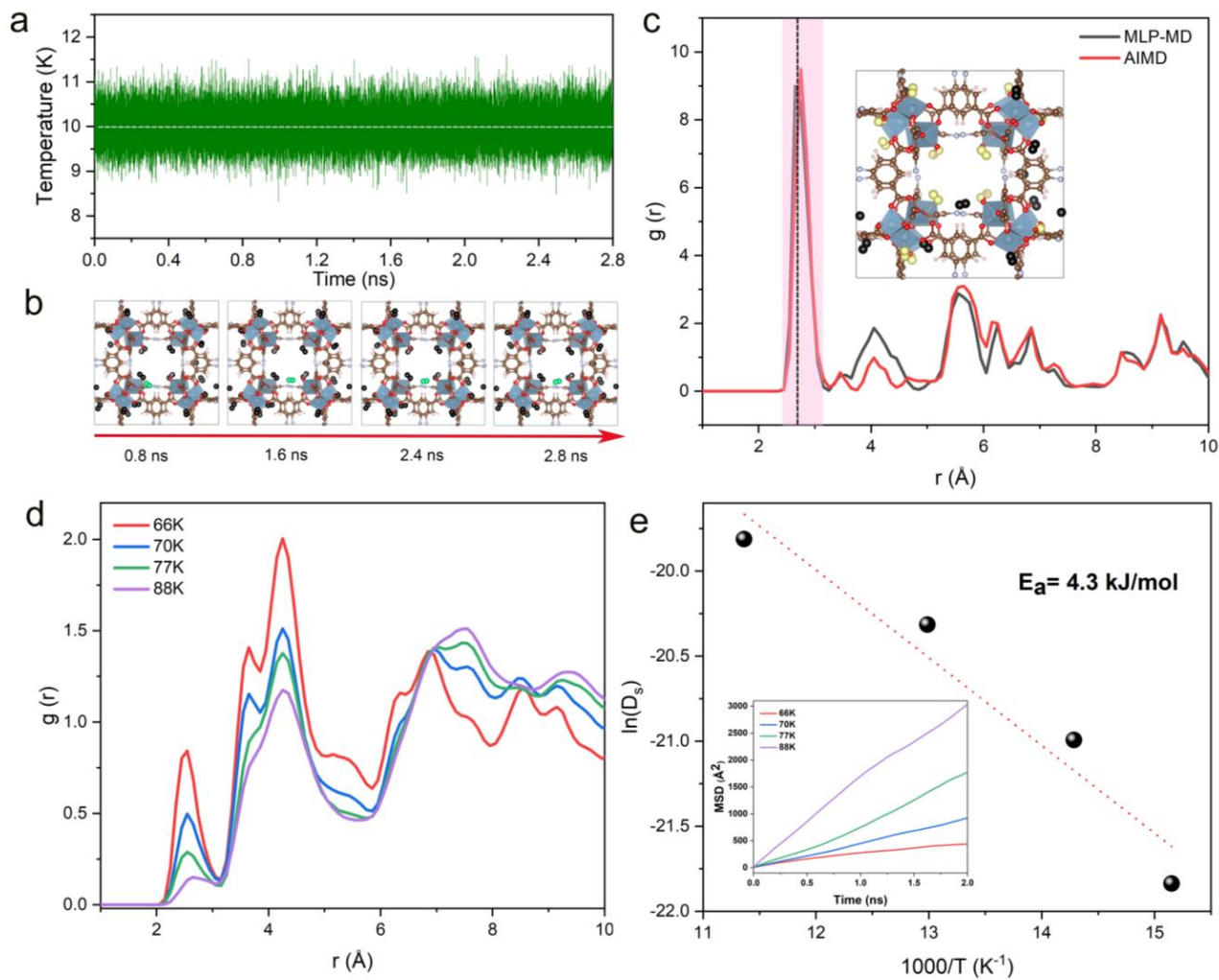

**Figure 3. MLP Molecular Dynamics Validation/Prediction for H$_2$@Al-soc-MOF.**
**(a) Temperature fluctuation of the MLP-MD for 24H$_2$@Al-soc-MOF at 10 K. (b) Illustrative snapshots of H$_2$@Al-soc-MOF at different MLP-MD times. H$_2$ molecules are depicted in black spheres, the color code for the rest of the MOF atoms is the same as in Figure 2 (except the H$_2$ molecules, which is in black here). To aid in tracking the motion of a specific H$_2$ molecule, one molecule is colored in green. (c) Radial distribution functions (RDFs) calculated for the Al-OMS/H$_2$ pair by MLP MD (black) and AIMD (red) both performed at 10 K. The dashed line, located at approximately 2.71 Å, corresponds to the equilibrium Al-H$_2$ distance obtained in the DFT-geometry optimized structure (0 K). In the insert, the yellow spheres represent H$_2$ molecules positioned at a distance of approximately 2.73 Å from the Al-OMS sites. (d) RDFs calculated for the Al-OMS/H$_2$ pair by MLP MD conducted at various temperatures (66, 70, 77, and 88K). (e) Arrhenius plot of the self-diffusion for H$_2$ simulated by MLP MD simulations. The insert shows the mean-squared displacement (MSD) for H$_2$ as a function of time calculated by MLP simulations performed at different temperatures.**

## MLP Molecular Dynamics/Monte Carlo Prediction

Long-scale MLP-MD simulations were thus performed to explore the dynamics of $H_2$ in Al-soc-MOF at different temperatures. The insert of Figure 4e shows that the resulting mean squared displacement (MSD) follows a linear time-dependence, signature of a Fickian diffusion regime. The self-diffusion coefficient ($D_s$) for $H_2$ was thus calculated for all temperature using the Einstein-relation. Herein $D_s$ of $H_2$ at 77 K is $1.9 \times 10^{-9}$ $m^2$/s which is slower than the value previously reported for MOFs free of OMS sites, like MIL-53, MIL-47, and IRMOF,[38, 50, 51] with $D_s$ ranging from $5\times10^{-9}$ to $5\times10^{-8}$ $m^2$/s at 77 K. Figure 4e reports the Arrhenius plot for $D_s$ associated to an activation energy of 4.3 kJ/mol. This energetic value is substantiality higher than that obtained for the MOFs free of OMS sites, i.e., MIL-53 (1.25 kJ/mol) and MIL-47 (0.68 kJ/mol),[38] IRMOF-1 (2.55 kJ/mol), IRMOF-8 (2.10 kJ/mol) and IRMOF-18 (3.09 kJ/mol).[50] while it is falls within the same range of value that was observed for zeolite NaX (4.0 kJ/mol) where $H_2$ can also interact with the extra-framework cations Na+.[52] Both slower self-diffusion coefficient and higher activation energy predicted for Al-soc-MOF compared to other MOFs mentioned above are in line with the strong interactions between $H_2$ and Al-OMS that tend to retain more $H_2$ localized around the inorganic node and indeed slow down its diffusivity.

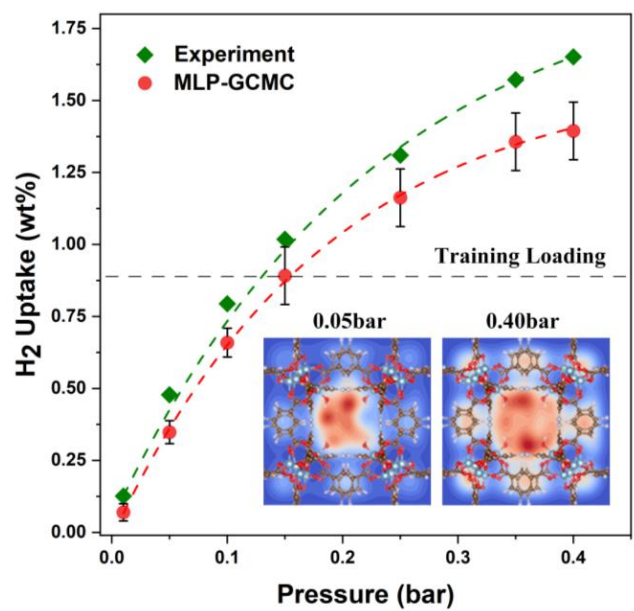

**Figure 4. MLP GCMC Prediction.** Simulated $H_2$ Adsorption isotherms in Al-soc-MOF (red dots) vs experimental gravimetric measurements (green pentagrams) at 77K, ranging from 0.01 to 0.4 bar. For reference, the loading used for training (24 $H_2$ molecules per unit cell) is depicted by a dashed line. The insert figures represent the probability distribution of $H_2$ at 0.05 and 0.4 bar.

Finally, the MLP was implemented in a grand-canonical Monte Carlo (GCMC) scheme to predict the adsorption isotherm of $H_2$ in Al-soc-MOF at 77 K. As stated above, this is the biggest challenge for the MOF community working in the field of adsorption to accurately predict the adsorption isotherm at very low pressure for a MOF containing OMS.[26] Herein the MLP GCMC simulations were performed at 77 K for pressure ranging from 0.01 bar to 0.4 bar, that surrounds the pressure used for the derivation of the MLP (0.15 bar). Note that all the simulations were performed starting with a 24 $H_2$@Al-soc-MOF configuration corresponding the $H_2$ loading used to derive the MLP and the gas pressure was subsequently applied to determine the amount adsorbed at equilibrium for each pressure point explored in the adsorption isotherm. This procedure is required since if we start with an empty Al-soc-MOF configuration, no $H_2$ molecule can be adsorbed because our MLP was derived to account for $H_2$-$H_2$ interactions, which are absent when there are no $H_2$ molecules initially in the system. The MLP GCMC simulated adsorption isotherm is reported in Figure 4.

To gain more insights into the adsorption mechanism, we analyzed the probability distribution of $H_2$ at 0.05 bar and 0.4 bar, as shown in the insert of Figure 4. In these snapshots, the red and bule regions are associated to high and low probability to find $H_2$ respectively. At very low pressure (0.05 bar), these plots confirm that $H_2$ molecules preferentially adsorb in the vicinity of the Al-OMS. At higher pressure (0.4 bar), more $H_2$ molecules are adsorbed, and these additional molecules occupy less energetically favorable adsorption sites in the pores resulting in a more homogeneous adsorption pattern. To validate these GCMC simulations, the Al-soc-MOF sample was synthesized and fully activated prior collecting its gravimetric $H_2$ adsorption isotherm (see ESI for details Figure S3). The overall good agreement obtained between the MLP-GCMC calculated and the experimental $H_2$ adsorption isotherms provide strong confirmation that the derived MLP accurately describe the interactions between $H_2$ and Al-OMS that dominate the initial stage of adsorption. This conclusion is also supported by the simulated adsorption enthalpy at low coverage of -6 kJ/mol which in line with the experimental value reported previously for the In-soc-MOF (-6.5 kJ/mol).[45] One can note that as the pressure increases, the deviation between experiment and simulation slightly increases. This is most probably attributed to the contribution of the adsorbate-adsorbate interactions which are less well described by our MLP when the adsorbed amount of $H_2$ exceeds the loading used for the derivation of the MLP.[53] These interactions play a more significant role at the higher pressure, affecting the adsorption behavior of molecules.

## CONCLUSIONS

In summary, we have derived a MLP trained from a set of trajectories generated by *Ab initio* Molecular Dynamics simulations to accurately describe the interaction between $H_2$ and Al-soc-MOF containing Al-OMS. A preliminary validation in terms of $H_2$ binding mode and temperature dependent $H_2$ distribution, was achieved by means MLP-MD simulations performed for temperature ranging from 10K to 88K. MLP-MD and MLP-GCMC simulations were further conducted at 77 K to gain an unprecedented microscopic insight into the thermodynamics adsorption and dynamics of $H_2$ in this MOF. This high-accuracy molecular simulation approach was validated by a good agreement between the predicted MLP-GCMC $H_2$ adsorption isotherm and the corresponding experimental data collected on a well-activated Al-soc-MOF sample. Decisively, such derived MLP overcomes the limitations of generic force fields that inaccurately the interactions between small molecules and MOFs with OMS that has hampered so far accurate prediction on MOFs with OMS for diverse adsorption-related applications. This first MLP related to $H_2$@MOF-OMS is of utmost importance since $H_2$ is known to be highly polarizable and exhibits quantum-mechanical effects at cryogenic temperature that are only accurately described by quantum mechanics. Beyond gaining an in-depth understanding of the $H_2$/MOF-OMS interactions of importance to further develop advanced MOFs for efficient $H_2$ storage, this computational strategy is expected to be more systematically applied for predicting the adsorption properties of many existing MOFs-OMS and made-to-order novel ones.

## METHODS

**DFT calculations.** A unit cell of Al-soc-MOF with the lattice parameter *a*=*b*=*c*=21.67 Å and *α*=*β*=*γ*=90° was loaded with 24 $H_2$ molecules corresponding to a 0.92%wt uptake. All the training and validation data-sets were generated by AIMD using the Vienna Ab-initio Simulation Package (VASP) with projector augmented wave (PAW) pseudopotential.[54, 55] The generalized gradient approximation (GGA) with Perdew-Burk-Ernzerhof (PBE) functionals was used to describe the exchange-correlation interaction of the electrons.[56] To account for the non-local, long-range electron corrections, Grimme empirical dispersion corrections (DFT-D3 method) were adopted in all DFT calculations.[57] The Kohn-Sham orbitals (wave functions) were expanded in a plane-wave basis set with a cut-off energy of 480 eV. The Brillouin zones was limited to the Γ point. For geometry optimization at 0K, the overall MOF/$H_2$ system was fully relaxed until both the energy and force reach the convergence criteria of $10^{-5}$ eV and 0.02 eV/Å, respectively. This geometry optimized configuration served as the initial stage for the AIMD simulations. All AIMD simulations were performed in the canonical ensemble (NVT) with the Nosé-Hoover thermostat to control the temperature at 77K for 7 ps. A timestep of 0.5 fs was employed to integrate the equations of motion.[58] Detailed information and corresponding data can be found in the supporting information, including a separate dataset folder containing all relevant data.

**MLP calculations**. The DeepMD-kit package was employed to train MLP from the configurations generated by AIMD. The DeepMD-kit framework, proposed by Zhang et al.,[47, 48] based on a deep neural network (DNN) learning method, has been shown to be highly effective in various MD simulation studies involving liquid, bulk materials, organic molecules, and solids.[59-61] In the DeepMD-kit approach, the potential energy of each atomic configuration is computed as the sum of the individual atomic energies:

$$E = \sum_i (E_i)$$

where $E_i$ is determined by the local environment of atom i within its near neighbors by setting a cutoff radius of $R_c$. To guarantee the preservation of the translational, rotational, and permutational symmetries of the potential energy surface, DeepMD-kit constructs a local structure descriptor {$D_j$} (local environment matrix) from Cartesian coordinates {$R_j$} of the neighboring j atom within the cutoff radius with respect to atom i. This construction is achieved through an embed neural network (NN) as the filter NN. Subsequently, the output of filter NN, which consists of the local descriptors, is mapped to a local atomic energy $E_i$ which is fully determined by the i atom and its near neighbors within the radius of $R_c$ by employing another end-to-end neural network called the fitting NN. The fitting NN is designed to be smooth and continuous, enabling a seamless transformation from local descriptors to the local atomic energy. The training process involves optimizing the parameters in both the filter and fitting NNs to minimize the loss function:

$$L = \frac{P_e}{N}|\Delta E|^2 + \frac{P_f}{3N}\sum_i |\Delta f_i|^2$$

where $|\Delta E|^2$ and $|\Delta f_i|^2$ represent the root mean square errors (RMSEs) of the energy and force, respectively. These errors quantify the discrepancies between the predicted values and the reference values obtained from the AIMD calculations. $P_e$ and $P_f$ act as perfectors during the training process, which are functions of the dynamic learning rate and continuously adapt during the optimization of the NNs.

In this work, the cutoff radius for neighbor searching was set to 7.0 Å, with a smooth cutoff of 4.1 Å. This choice was motivated by the pore size/shape of Al-soc-MOF, which requires considering interactions beyond the immediate vicinity of an atom to capture its behavior accurately. The maximum number of neighbors within the cutoff was set to 70 based on configuration. The size of the filter and fitting NNs were chosen as {25, 50, 100} and {240, 240, 240}, respectively. To control the training process, the decay rate and decay step parameters were set to 0.95 and 5000. The initial values of perfectors $P_e$ and $P_f$ in the loss function were set to 0.02 and 1000. For training, the Adam stochastic gradient descent method was adopted, which is an efficient algorithm for optimizing neural networks.[62] The learning rate starts at 0.001 and decreases exponentially over the training process to facilitate the convergence of the MLP.

**MLP-MD simulations.** The trained-MLP was implemented into a MD scheme with the MLP executed using the DeepMD-kit interface to the LAMMPS code.[63] These MD simulations were carried out in the NVT thermostat ensemble using a timestep of 0.5 fs, and the duration of the MD simulations is at the nanosecond (ns) level.

**MLP-GCMC simulations.** The trained-MLP was implemented into a MC scheme using an in-house developed MC code in conjunction with LAMMPS program.[63] For the GCMC simulations performed at 77K, a series of random moves for $H_2$ including insertion (30%), deletion (30%), translation (20%), and rotation (20%) were considered. The acceptance or rejection of these moves was determined by calculating the Boltzmann probability of different configurations. Additionally, the potential energy functions for different positions and orientations of gas molecules were considered to determine the probability density of gas molecules at various adsorption sites.

**MOF Synthesis and characterization.** A mixture of $AlCl_3 \cdot 6H_2O$ (13mg, 0.054 mmol) and 3,30,5,50 – azobenzene-tetracarboxylic acid (10 mg, 0.028 mmol) was dissolved in N,N-dimethylformamide (DMF) (2mL), acetonitrile ($CH_3CN$) (2 mL), and acetic acid (1mL). The solution was carefully transferred into Pyrex vial with phenolic cap lined with polytetrafluoroethylene (PTFE). The vial was then placed in a preheated oven at 150°C for duration of 3 days, resulting in the formation of pure orange crystals.

The Al-soc-MOF was activated by first washing the synthesized crystals 3 times with DMF. Subsequently, the crystals were exchanged with acetonitrile for 6 days in a 65°C oven. The guest solvents in the pore system of Al-soc-MOF were fully removed by a traditional activation approach (vacuum and heating). The calculated apparent BET area and pore volume were 1534 $m^2/g$ and 0.6 $cm^3/g$.

Low pressure gas adsorption measurements for $H_2$ were performed on 3-Flex Surface Characterization Analyzer (Micromeritics) at relative pressures up to 1 bar. These experiments were carried at 77 K, for Al-soc-MOF after 240°C activation (Figure S3). The bath temperature for the $H_2$ sorption was controlled using liquid nitrogen bath at 77 K.

## AUTHOR INFORMATION

**Corresponding Author**

**Guillaume Maurin** – UMR 5253, CNRS, ENSCM, Institute Charles Gerhardt Montpellier, University of Montpellier, Montpellier 34293, France
E-mail: guillaume.maurin1@umontpellier.fr

**Authors**
**Shanping Liu** – UMR 5253, CNRS, ENSCM, Institute Charles Gerhardt Montpellier, University of Montpellier, Montpellier 34293, France
**Romain Dupuis** – LMGC, Univ. Montpellier, CNRS, Montpellier, France
**Dong Fan** – UMR 5253, CNRS, ENSCM, Institute Charles Gerhardt Montpellier, University of Montpellier, Montpellier 34293, France
**Salma Benzaria** – Division of Physical Science and Engineering, Advanced Membrane and Porous Materials Center, King Abdullah, University of Science and Technology (KAUST), Thuwal 23955-6900, Kingdom of Saudi Arabia
**Michael Bonneau** – Division of Physical Science and Engineering, Advanced Membrane and Porous Materials Center, King Abdullah, University of Science and Technology (KAUST), Thuwal 23955-6900, Kingdom of Saudi Arabia
**Prashant Bhatt** – Division of Physical Science and Engineering, Advanced Membrane and Porous Materials Center, King Abdullah, University of Science and Technology (KAUST), Thuwal 23955-6900, Kingdom of Saudi Arabia
**Mohamed Eddaoudi** – Division of Physical Science and Engineering, Advanced Membrane and Porous Materials Center, King Abdullah, University of Science and Technology (KAUST), Thuwal 23955-6900, Kingdom of Saudi Arabia

**Notes**
The authors declare no competing financial interest.

## ACKNOWLEDGMENT
The computational work was performed using HPC resources from GENCI-CINES (Grant A0140907613).

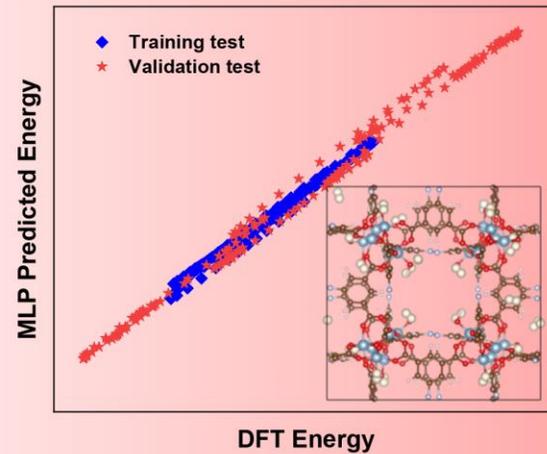

TOC